**RESEARCH ARTICLE**

# Performance Benchmarking of Psychomotor Skills Using Wearable Devices: An Application in Sport

MAHELA PANDUKABHAYA[1,*], (Member, IEEE), THARAKA FONSEKA[1,*],
MADHUMINI KULATHUNGE[1], (Member, IEEE),
ROSHAN GODALIYADDA[1], (Senior Member, IEEE),
PARAKRAMA EKANAYAKE[1], (Senior Member, IEEE),
CHANAKA SENANAYAKE[2], AND VIJITHA HERATH[1], (Senior Member, IEEE)

[1]Department of Electrical and Electronic Engineering, Faculty of Engineering, University of Peradeniya, Peradeniya 20400, Sri Lanka
[2]Department of Manufacturing and Industrial Engineering, Faculty of Engineering, University of Peradeniya, Peradeniya 20400, Sri Lanka

Corresponding author: Mahela Pandukabhaya (e17234@eng.pdn.ac.lk)

*Mahela Pandukabhaya and Tharaka Fonseka are co-first authors.

**ABSTRACT** Mastering psychomotor skills, such as those essential in sports, rehabilitation, and professional training, often requires a precise understanding of motion patterns and performance metrics. This study proposes a versatile framework for optimizing psychomotor learning through human motion analysis. Utilizing a wearable IMU sensor system, the motion trajectories of a given psychomotor task are acquired and then linked to points in a performance space using a predefined set of quality metrics specific to the psychomotor skill. This enables the identification of a benchmark cluster in the performance space, which represents a group of reference points that define optimal performance across multiple criteria, allowing correspondences to be established between the performance clusters and sets of trajectories in the motion space. As a result, common or specific deviations in the performance space can be identified, enabling remedial actions in the motion space to optimize performance. A thorough validation of the proposed framework is done in this paper using a Table Tennis forehand stroke as a case study. The resulting quantitative and visual representation of performance empowers individuals to optimize their skills and achieve peak performance.

## I. INTRODUCTION

Human motion analysis is a crucial area of study with wide-ranging relevance and broad applications across numerous disciplines including, but not limited to, biomechanics, computer vision, robotics, healthcare, and sports science [1], [2], [3], [4]. By meticulous analysis of human movement patterns, such as gait, posture, and hand gestures, researchers can gain insights into the underlying dynamics and functions of the human body. These insights are vital for developing advanced prosthetics, improving athletic performance, enhancing rehabilitation protocols, and driving innovations in human-machine interactions [5], [6], [7], [8].

Wearable technology has emerged as a powerful tool for human motion analysis [9], enabling prolonged monitoring of physical activities in real-world settings. Recent developments in data analytics, including AI, coupled with innovations in measurement devices for motion tracking, have made it possible to acquire valuable insights into motion analysis while causing minimal disruption to the

The associate editor coordinating the review of this manuscript and approving it for publication was Razi Iqbal.

underlying motion profile or activity [10]. These technologies allow for continuous monitoring of physical activities [11] and offer valuable insights for applications such as sports performance optimization, healthcare, workplace ergonomics to prevent injuries, military training to track physical readiness, and in Virtual Reality (VR) to enhance immersive experience by accurately capturing user movements [11], [12], [13], [14]. Furthermore, wearable technologies offer several advantages over traditional motion capture systems such as portability, affordability, and expandability [12], [15]. Concerns about privacy and identity protection can be easily addressed in wearable technologies as opposed to video-based systems [16], [17].

Numerous wearable sensor options exist, such as inertial measurement unit (IMU) sensors, Electromyography (EMG) sensors, force transducers, and pressure sensors for monitoring various aspects of human motion and physiology [18], [19]. Each sensor type captures specific data, enabling detailed insights into an individual's physical state or movement patterns. For example, EMG sensors monitor muscle activity, Electrocardiography (ECG) sensors track heart health, force transducers evaluate applied forces, and pressure sensors measure force distribution across surfaces [11].

Sensors such as EMG, ECG, force transducers, and pressure sensors provide valuable physiological data, but they often require direct skin contact, calibration, or specialized setups [11], [20]. Such complexities make them impractical for motion analysis and are less commonly used for capturing comprehensive movement data. In contrast, IMU sensors, in particular, are widely used in wearable technology for motion analysis [21]. They combine accelerometers, gyroscopes, and magnetometers to capture precise movement data, including acceleration, rotation, and orientation of the body in space [21]. One of the key advantages of IMUs is that they only capture signal patterns specific to an individual's movements, significantly reducing the risk of compromising their privacy. Furthermore, IMUs play a crucial role in enhancing the capabilities of wearable technology, allowing for the capture of subtle movements that traditional systems may overlook [22]. Additionally, they enable the development of low-cost, non-invasive wearable devices that cater to numerous applications in prosthetic grading, vocational training, and sports skills monitoring [23], [24], [25].

The connection between human motion analysis and psychomotor learning is profound. By understanding movement patterns, researchers can gain valuable insights into how individuals acquire and refine motor skills [26], [27], [28]. Psychomotor learning involves the coordination between cognitive functions and physical movements, and analyzing these motions allows for a deeper comprehension of how individuals acquire and improve skills such as coordination, balance, and precision. By examining movement patterns, it is possible to identify areas where individuals struggle, enabling targeted interventions to enhance learning outcomes. This is particularly important in areas like sports [29], [30], rehabilitation [31], industrial tasks [32] and skill-based education [33], [34], where mastery of physical tasks is essential. Due to the inherent subjectivity and other challenges of psychomotor skill learning, a quantitative data-driven approach for analyzing human motion during psychomotor tasks would be advantageous to optimize training methods and promote more effective learning processes.

In this study, we introduce a framework for optimizing psychomotor learning through human motion analysis. By utilizing an in-house fabricated wearable device equipped with IMU sensors [32], we capture individual motion patterns when a psychomotor task is executed. The collected motion data are then mapped into a space called performance space, where unsupervised clustering techniques are applied to identify a benchmark cluster, allowing us to recognize the best executions within the sample population. Here, a specific movement trajectory of the designated task in the motion space is represented by a data point in the performance space. Hence the data-driven approach allows clusters to be formed based on the performance relative to the quality indices, enabling the observation of ideal and suboptimal executions for a given task in the motion space. This also enables trackability in the performance space, facilitating not only the identification of the benchmark but also the common and specific deviations from the benchmark.

For any psychomotor task, specific parameters exist that define optimal execution. For instance, certain techniques are recommended in sports to improve performance, and training could be guided by these parameters indicating the ideal performance of a given sports technique. Similarly, precise hand movements are crucial in surgical procedures, and training could be enhanced by defining parameters that practitioners need to excel in. These parameters could be derived from various ways, including product outcomes in industrial processes, live video analysis of movement patterns, or any other suitable method specific to the psychomotor task that is being considered.

We demonstrate the effectiveness of the framework proposed in this paper through a detailed case study of table tennis. By analyzing the movement patterns of table tennis players executing a forehand stroke, we can identify optimal techniques and provide personalized feedback to improve performance. This approach is not confined to this sport; by introducing task-specific parameters, the framework can be easily adapted to various other psychomotor learning contexts, including sports, healthcare, and industrial training environments [32].

Given the research gaps identified in the existing literature on recent work in the field, the contributions of this study can be listed as follows.

- *Motion space to performance space mapping:* Introducing a framework that binds 3D real-world motion data of a given psychomotor task with a performance space generated through criteria set forth related to the task quality.
- *Identification of the benchmark and the other clusters:* Proposing a data-driven unsupervised clustering method

that enables identifying the number of clusters in the performance space including the benchmark and the deviating clusters.
- *Interpretation for the divergence of the clusters from the benchmark in the motion space:* Identifying the common deviations of the clusters from the benchmark and re-interpreting them in the motion space to get valuable insights on what type of motion discrepancies have led to such performances.
- *Verification of the methodology using a comprehensive case study:* Validating the aforementioned generalized framework using a real-world example, i.e., table tennis with a predefined set of performance parameters, visualizing and interpreting the results.

In Section II, the proposed methodology is explained in detail, and in Section III, the applicability of this technique through a case study that involves the assessment of a psychomotor skill, specifically, a forehand stroke in table tennis is demonstrated.

## II. PROPOSED METHODOLOGY

In this section, we elaborate on the formal methods incorporated into the selection of a psychomotor skill, data collection using the wearable device, data preprocessing, use of extended Kalman filter (EKF) to create the Euler angle space, deriving the performance space from domain expertise along with literature, and clustering this space to identify the benchmark and deviating clusters. An overview of this proposed framework is presented in Fig. 1.

### A. PRELIMINARIES
#### 1) SELECTION OF A PSYCHOMOTOR SKILL
Firstly, an appropriate psychomotor skill should be selected for the experiment, one that has specific physical variables that can be extracted. These variables imply the quality of the executed task as well as the relevant techniques that need improvement for the overall performance. There can be a vast scope of such tasks in different fields. Hence, a suitable task must be chosen based on the specific requirements.

#### 2) SELECTION OF PSYCHOMOTOR SKILL-SPECIFIC QUALITY PARAMETERS
After the selection of the psychomotor skill, it is required to select a set of parameters suited to evaluate the task. This set of parameters should reflect the relevant techniques a person must excel in order to excel in the given psychomotor skill.

### B. DATA COLLECTION AND PREPROCESSING
#### 1) ETHICAL CLEARANCE
The commitment of the project to ethical research practices was underscored by obtaining ethical clearance from the Faculty Ethics Review Committee (reference no.: ERC/FoE/2022/003).

Prior to data collection, consent forms were provided to participants, collected, and documented as part of the ethical clearance process under the mentioned reference number. Additionally, data protection measures were implemented in accordance with the guidelines specified in the same ethical clearance documentation. This ethical diligence ensured the integrity of data collection and analysis while fostering credibility and ensuring adherence to ethical standards in the research.

**TABLE 1.** Specifications of the wireless wearable device.

| Symbol | Parameter | Value |
|---|---|---|
| $F_s$ | Data sampling rate | 64 Hz |
| $N'_m$ | Number of modules | 16 |
| $m_i$ | Weight per module | $\approx 40$ g |
| $V_w$ | Working voltage of a module | 3.3 V |

#### 2) WIRELESS WEARABLE DEVICE
A wireless wearable device was developed by the authors [32] allowing for smooth and parallel data collection and improving the overall research framework with the integration of Inertial Measurement Units (IMU) sensors. The device, which is shown in Fig. 1 (a) is mainly composed of $N'_m = 16$ such IMUs, specifically MPU6050 sensors, which can be worn on both the hands and legs of a person.

Although the full-body device contains 16 IMUs, a subset of these sensors can be selected for a given psychomotor task by removing excess sensors unrelated to the action, mitigating their impact on the observations. This allows for a flexible selection of the sensors around the joints that are most effective in the evaluation of the selected psychomotor skill [35]. In addition, the device was made more lightweight minimizing the device's impact on the person's performance. Hence, for this study, $N_m = 4$ sensors were used as the dominant arm was considered for the data collection procedure. The four modules were attached to the wrist, forearm, biceps, and shoulder as shown in Fig. 2. Furthermore, the device was validated using YOLOv7 [36], [37] which is a popular computer vision-based object detection platform.

The device consists of two types of modules namely: Alpha ($\alpha$), and Beta ($\beta$). A $\beta$ module mainly comprises an IMU sensor, and an SD card, connected to an ESP-32S microcontroller and powered by a 1200 mAh, 3.7 V LiPo battery. An $\alpha$ module is similar to a $\beta$ module in terms of components, but lacks an IMU sensor. The $\alpha$ module hosts a web server within its SD card and uses its logic to control the $\beta$ modules. Based on the control instructions sent from the $\alpha$ module via the in-built ESP-NOW protocol, a $\beta$ module samples the data from its IMU, and writes the acquired data to its SD card as a comma-separated values (CSV) file for post-experimental retrieval. The additional specifications of the device are listed in Table 1.

The accelerometric data $[a_x, a_y, a_z]$ represents the variation of acceleration of the attached limb, including gravitational acceleration. On the other hand, the gyroscopic data $[g_x, g_y, g_z]$ represents the angular velocity vector of the limb to which it is attached. Collectively, an MPU6050 sensor outputs a vector $[a_x, a_y, a_z, g_x, g_y, g_z]$, which accounts for $N_X = 6$ degrees of freedom per sensor. As there are $N_m = 4$ modules, the total number of dimensions of motion data comes as $N_{X,\text{IMU}} = N_X \times N_m = 24$. This defines the IMU data space as $\mathbb{I} \equiv \mathbb{R}^{24} \times \mathbb{Z}_0^+$.

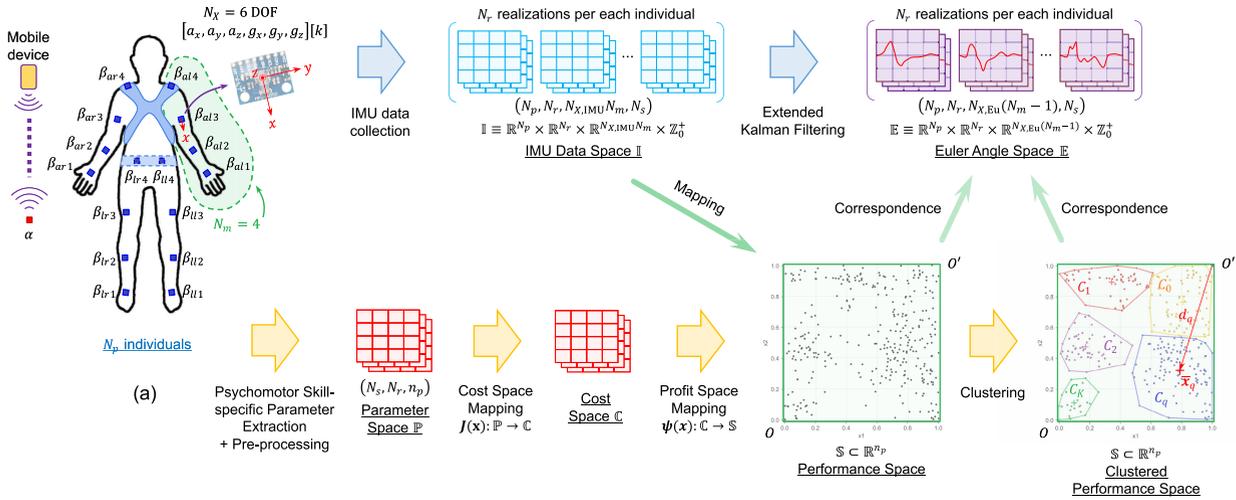

**FIGURE 1.** Overview of the framework.

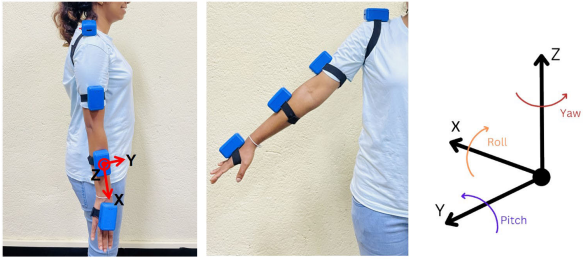

**FIGURE 2.** Sensor placement in the dominant arm.

The raw IMU data for each individual were split into realizations using a signal annotation tool developed by the authors, with each realization capturing one cycle of the signal waveform.

### C. COMPUTATION OF EULER ANGLES FROM IMU DATA: EKF

For better visualization of the obtained IMU data, Euler angles were incorporated as an alternative representation of IMU signals. Euler angles capture invaluable and visualizable information on the movement of limbs in a chosen psychomotor skill.

The most straightforward way to obtain joint angles from the IMU data is to integrate the sensor angular velocity of each limb. However, this approach is typically susceptible to gyroscope drift [38] along with sensor noise [39], rendering it unsuitable for joint angle estimation. To surmount this, a modified version of the discretized Extended Kalman Filter (EKF), which is commonly emphasized in literature [40], [41], was adopted.

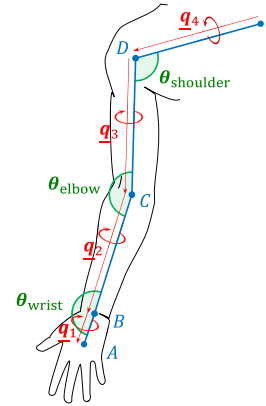

**FIGURE 3.** Euler angles in the arm.

The resulting IMU data space $\mathbb{I}$ was pre-filtered and fused using the equations (1)-(14), with notations provided in Table 2, to obtain the orientation quaternions $\mathbf{q}_1, \mathbf{q}_2, \mathbf{q}_3, \mathbf{q}_4$ of each limb that represents its axis and the angle of rotation as shown in Fig. 3. Quaternions were used to reduce the computational complexity that arises from using rotational matrices in the EKF framework. The obtained quaternions were used in pairs to estimate the Euler angles of joints $\theta_{\text{shoulder}}, \theta_{\text{elbow}}, \theta_{\text{wrist}} \in \mathbb{R}^3$ using quaternion multiplication. Overall, there are $N_{X,\text{Eu}} = 3 \times 3 = 9$ components of angles for a given time instance. Hence, this process results in the Euler angle space $\mathbb{E} \equiv \mathbb{R}^9 \times \mathbb{Z}_0^+$ for the limb, thus reducing the raw IMU data to Euler angles through $\mathbb{R}^{24} \times \mathbb{Z}_0^+ \to \mathbb{R}^9 \times \mathbb{Z}_0^+$, which forms a contraction mapping.

*State Equations:*

$$\mathbf{x}_k = \begin{bmatrix} \hat{\mathbf{q}}_k \\ \hat{\mathbf{b}}_k \end{bmatrix} \quad (1)$$

$$\tilde{\mathbf{x}}_k = \begin{bmatrix} \delta\theta_k \\ \Delta\hat{\mathbf{b}}_k \end{bmatrix} \quad (2)$$

**TABLE 2.** Notations for EKF framework.

| Notation | Description |
|---|---|
| $k$ | Time step |
| $x_k$ | State Vector |
| $\tilde{x}_k$ | Error Vector |
| $\hat{b}_k$ | Gyro Bias |
| $\delta\theta_k$ | Error Angle Vector |
| $\Delta\hat{b}_k$ | Gyro Bias Error |
| $\hat{\omega}_k$ | Estimated Rotational Velocity |
| $\omega_{m_k}$ | Measured Rotational Velocity |
| $P_{k|k-1}$ | State Covariance matrix |
| $F_k$ | State Transition Matrix |
| $Q_k$ | Noise Covariance matrix |
| $\tilde{r}_{k|k-1}$ | Measurement Error |
| $z_k$ | Actual Measurement |
| $\hat{z}_{k|k-1}$ | Measurement Estimation |
| $S_k$ | Residual Covariance Matrix |
| $R_k$ | Measurement Error Covariance |
| $H_k$ | Measurement Matrix |
| $K_k$ | Kalman Gain Matrix |
| $\Delta\hat{x}_k$ | State Correction Matrix |
| $\underline{\hat{q}}_{k|k}$ | Quaternion |

*Propagation Equations:*

$$\hat{b}_{k|k-1} = \hat{b}_{k-1|k-1} \quad (3)$$

$$\hat{\omega}_{k|k-1} = \omega_{m_k} - \hat{b}_{k|k-1} \quad (4)$$

$$P_{k|k-1} = F_k P_{k-1|k-1} F_k^T + Q_k \quad (5)$$

*Update Equations:*

$$\tilde{r}_{k|k-1} = z_k - \hat{z}_{k|k-1} \quad (6)$$

$$S_k = H_k P_{k|k-1} H_k^T + R_k \quad (7)$$

$$K_k = P_{k|k-1} H_k^T S_k^{-1} \quad (8)$$

$$\Delta\hat{x}_k = \begin{bmatrix} \delta\hat{\theta}_k \\ \Delta\hat{b}_k \end{bmatrix} = K_k \tilde{r}_{k|k-1} \quad (9)$$

$$2\delta\underline{\hat{q}}_k = \begin{bmatrix} \delta\hat{\theta}_k \\ 1 \end{bmatrix} \quad (10)$$

$$\hat{b}_{k|k} = \hat{b}_{k|k-1} + \Delta\hat{b}_k \quad (11)$$

$$\hat{\omega}_{k|k} = \omega_{m_k} - \hat{b}_{k|k} \quad (12)$$

$$\underline{\hat{q}}_{k|k} = \frac{\delta\underline{\hat{q}}_k}{\|\delta\underline{\hat{q}}_k\|} \otimes \underline{\hat{q}}_k \quad (13)$$

$$P_{k|k} = (I - K_k H_k) P_{k|k-1} \quad (14)$$

### D. EXTRACTION OF PSYCHOMOTOR SKILL-SPECIFIC PARAMETERS

The set of quality parameters defined earlier for the selected psychomotor task can be extracted through a practical methodology from various sources. For example, real-time video feeds offer a versatile method for capturing and analyzing dynamic processes. By using computer vision techniques, it is possible to extract instantaneous metrics, such as velocity, from recorded motion trajectories. Alternatively, quantitative parameters can be obtained from static analysis of end products. For instance, in industrial manufacturing, precise measurements of attributes like weld width can be extracted from physical specimens. If $n_p$ number of such parameters are extracted, a parameter space $\mathbb{P} \subset \mathbb{R}^{n_p}$ can be defined to obtain various performance levels of groups of people executing the relevant motion.

### E. THE PERFORMANCE SPACE

#### 1) PERFORMANCE SPACE AND COST FUNCTIONS

After the extraction of $n_p$ quality parameters, a performance space $\mathbb{S} \subset \mathbb{R}^{n_p}$, in particular, $\mathbb{S} \equiv [0, 1]^{n_p}$, is defined to assign a score to a selected action.

The first step is to define a set of cost functions $\mathbf{J}(\mathbf{x}) \equiv \left[J_1(x_1), \ldots, J_{n_p}(x_{n_p})\right]^T$ which assign a cost to each parameter that decreases as the quality of the parameter improves. The definition of these cost functions is based on domain expertise in the field of the chosen psychomotor skill.

#### 2) NORMALIZATION THROUGH RADIAL BASIS FUNCTIONS (RBFs)

As the output ranges of each cost function may vary depending on the specific parameter, it is essential to convert the costs to a common range so that they can be compared. To do so, we define a set of Gaussian radial basis functions (RBFs) $\psi(\mathbf{x}) \equiv \left[\psi_1(x_1), \ldots, \psi_{n_p}(x_{n_p})\right]^T$ to normalize the costs obtained through each cost function to the common range $[0, \psi_0]$:

$$\psi_p : \mathbb{R} \to [0, \psi_0], \ \psi_p(x_p) = \psi_0 e^{-\alpha_p^2 (x_p - \mu_p)^2}, \quad (15)$$

where $\alpha_p \in \mathbb{R}^+$ is a shape parameter defined for each parameter $p_1, \ldots, p_{n_p}$ based on the recommendations by field experts, and $\psi_0 \in \mathbb{R}^+$ is a constant which represents the upper bound of the transformation.

This transformation inverts the costs, converting them into 'profits' or performance scores. The entire parameter space $\mathbb{P} \subset \mathbb{R}^{n_p}$ is thus transformed to a performance space $\mathbb{S} \subset \mathbb{R}^{n_p}$.

### F. CLUSTERING OF THE PERFORMANCE SPACE

After obtaining the performance space $\mathbb{S}$, clustering is utilized to identify any inherent partitions within the performance space. This approach contrasts with manually categorizing the data, to use a data-driven approach thereby introducing a generalizability to our methodology. For this purpose, two main clustering methods are employed: spectral clustering, to identify the number of clusters, and $k$-means clustering to cluster the performance space $\mathbb{S}$.

#### 1) SPECTRAL CLUSTERING

Spectral clustering is a machine learning technique commonly used when data-driven approaches are required, especially for identifying clusters in data that may not be linearly separable. As shown in the literature, spectral clustering has been a successful technique for determining clusters in complex datasets, demonstrating reliable performance in various applications where traditional clustering methods may fall short [42], [43], [44]. Therefore, the largest single cluster in the high-dimensional space, formed by merging several sub-clusters, can be effectively clustered using this technique. As we zoom in on the cluster space, sub-clusters start to form within the feature space. It is possible to detect a number of small clusters submerged inside the initial supercluster at different levels of zooming. This phenomenon is known as "modes of clustering" [45], [46], [47]. Additionally, the number of clusters in a mode serves as the unique identifier for that mode. Therefore, different modes are identified by adjusting the free parameter $\sigma$, which represents the zooming effect in the standard spectral clustering algorithm. This procedure is referred to as "sigma sweep". The primary steps of the sigma sweep are as follows:

1) Generate the Affinity matrix $A$ for the dataset using the transformation:

$$A_{i,j} = \begin{cases} \exp\left(-\frac{\|x_i - x_j\|^2}{2\sigma^2}\right), & \text{for } i \neq j \\ 0, & \text{for } i = j \end{cases} \quad (16)$$

where $\sigma$ is a tunable parameter.

2) Compute the degree matrix $D$. $D$ is a diagonal matrix where element $(i, i)$ of $D$ is the row sum of $A$ in $i^{\text{th}}$ row:

$$D_{i,k} = \begin{cases} \sum_j A_{i,j}, & \text{for } i = k \\ 0, & \text{otherwise} \end{cases} \quad (17)$$

3) Calculate the Laplacian matrix $L$:

$$L = I - D^{-1/2} A D^{-1/2} \quad (18)$$

where $I$ is an identity matrix.

4) Obtain eigenvalues of $L$ and arrange them in descending order. Then compute eigengaps by obtaining the difference between two successive eigenvalues.

5) Plot the graph of the variation of the eigengap with $\log(\sigma)$ by repeating steps 1-4 for different $\sigma$ values to determine the number of sub-clusters $K$.

The characteristics of the sigma sweep are utilized to determine the ideal number of clusters for the performance space $\mathbb{S}$, $K_{\text{sc}}$. In essence, the required number of clusters is indicated by the eigengap curve that predominates over a wide range of $\sigma$ values.

Once the number of clusters $k$ is determined using the spectral clustering algorithm, $k$-means clustering is used to cluster the population using the determined $k$ value.

2) K-MEANS CLUSTERING

The $k$-means algorithm is an unsupervised algorithm that groups similar data points into $K$ clusters. It aims to minimize the distance between data points within each cluster. This is achieved by iteratively assigning data points to the nearest centroid and updating the centroids until convergence [48], [49], [50], [51].

In this study, $k$-means clustering is applied to the performance space $\mathbb{S}$ to obtain clusters $C_0, \ldots, C_{K_{\text{sc}}-1}$ after determining the number of clusters ($K_{\text{sc}}$) through the spectral clustering algorithm.

G. IDENTIFICATION OF THE BENCHMARK CLUSTER

A point can be defined in the performance space $\mathbb{S}$ that represents the ideal performance. Intuitively, this corresponds to the coordinate whose components are equal to the highest score, which is unity. We define this point as the origin of ideal performance $O' \equiv (1, \ldots, 1)$, which is not practically attainable. However, it is still possible to reach the neighborhood of $O'$. Let $\mathbf{x}_{O'}$ be the position vector of $O'$ with respect to the origin $O$ of $\mathbb{S}$.

We then identify the benchmark cluster $C_b$, which is anticipated to comprise the actions with the highest performance scores, through a distance criterion on the clusters defined through their relative placement in $\mathbb{S}$. Firstly, the centroid $G_q$ of all the $n_q$ points in each cluster $q = 0, \ldots, K_{\text{sc}} - 1$ is determined by:

$$\bar{\mathbf{x}}_q = \frac{1}{n_q} \sum_{j=0}^{n_q} \mathbf{x}_j. \quad (19)$$

Then, the $n_p$-dimensional Euclidean distance $d_q$ (2)-norm) of each centroid $G_q$ from $O'$ is computed by using:

$$d_q = \|\bar{\mathbf{x}}_q - \mathbf{x}_{O'}\|_2 = \sqrt{\sum_{j=1}^{n_p} (\bar{x}_{q,j} - x_{O',j})^2} \quad (20)$$

Finally, the benchmark cluster is defined as the cluster $C_b$ where:

$$b = \underset{q=0,\ldots,K_{\text{sc}}-1}{\arg\min}(d_q), \quad (21)$$

thereby choosing the cluster closest to $O'$.

H. IDENTIFICATION OF THE MEAN EULER SIGNAL

To capture the average behavior of Euler angles within each cluster, the mean Euler signal $\bar{\mathbf{Y}}_w[k] \in \mathbb{R}^9 \times \mathbb{Z}_0^+$ is computed by taking the ensemble average of all the Euler signal realizations $\mathbf{Y}_w[k; j]$ of axes $w = 1, \ldots, 9$ lying within that cluster:

$$\bar{\mathbf{Y}}_w[k] = \frac{1}{n_q} \sum_{j=1}^{n_q} \mathbf{Y}_w[k; j], \quad \forall w \in [1, \ldots, 9]. \quad (22)$$

where $n_q$ is the number of realizations within the considered cluster.

I. CORRESPONDENCE BETWEEN THE EULER ANGLES AND THE PERFORMANCE SPACE

As the final step, the Euler angle space $\mathbb{E} \equiv \mathbb{R}^9 \times \mathbb{Z}_0^+$ and the clustered performance space $\mathbb{S}' \equiv [0, 1]^{n_p}$ are compared qualitatively to observe the correspondence and the mapping between the two spaces. The connection from $\mathbb{E}$ to $\mathbb{S}'$ can be considered as contraction mapping, while the converse can be seen as correspondence. The deviations of the other clusters from the benchmark in the performance space can be explained by considering the Euler angle variations where the discrepancies in angles in the motion space $\mathbb{E}$ manifest as degradations of the task quality in the performance space. This can be identified as a breakdown analysis of the motion with the use of joint angles since it is possible to track which errors in the task execution have influenced performance outcomes.

## III. CASE STUDY: TABLE TENNIS

To demonstrate the applicability of our proposed methodology, a real-world case study on a psychomotor skill was used: a forehand stroke of table tennis.

### A. SELECTION OF THE PSYCHOMOTOR SKILL

Table tennis serves as an excellent example of both a competitive sport and a complex psychomotor task [52], requiring a combination of agility, coordination, and precision. Although table tennis may not be a dominant sport in numerous countries, it is estimated that approximately 300 million people worldwide [53] engage in this sport. Out of this number, at least 40 million are active competitive players [54].

This study focuses on the forehand stroke [55], as it is one of the most fundamental and widely used techniques due to its simplicity and repeatability [56]. Also, it is a technique that provides players with better control over the ball, enabling them to maintain consistent trajectories and placement. The stroke involves hitting the ball with the racket on the dominant side of the player's body, typically using a forward and slightly upward motion to achieve both speed and accuracy.

### B. DEFINITION OF PERFORMANCE PARAMETERS

As previously mentioned, this framework relies on the selection of parameters that are relevant to the selected psychomotor task, in our case, a forehand stroke in table tennis. Pertaining to this, we identified five such physically extractable quality parameters ($n_p = 5$) that characterize a successful forehand stroke:

- Bounce Distance $X$ ($p_1$)
- Bounce Distance $Y$ ($p_2$)
- Net Clearance ($p_3$)
- Instantaneous Ball Speed ($p_4$)
- Height Ratio ($p_5$)

It is important to emphasize that these five parameters are not exhaustive; researchers can select parameters tailored to their specific psychomotor task requirements. To illustrate our framework and the underlying principles of our methodology, we begin with these five parameters. In the subsequent paragraphs, we will explain the rationale behind selecting these parameters, along with the technical background that informed our choices. We consulted experienced coaches and players, as well as relevant literature, to guide our selection process.

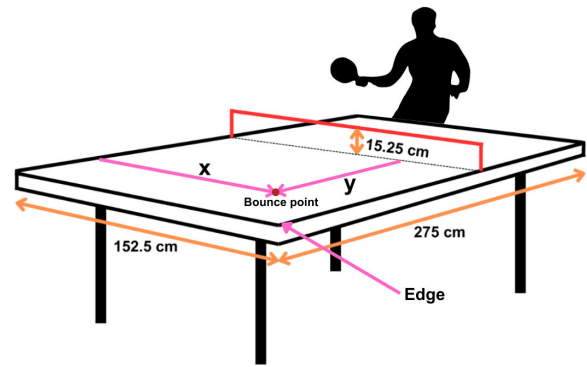

**FIGURE 4.** Parameters of table tennis (table width, table height, and net height) along with our definitions for bounce distances X, and Y for a right-handed player.

#### 1) BOUNCE DISTANCE X AND BOUNCE DISTANCE Y

When executing a forehand stroke in table tennis, it is recommended for the player to aim for the ball to land as far from the net as possible, while still bouncing within the table. Landing the ball closer to the edge makes it harder for the opponent to return [57]. A stroke placed near the edge significantly reduces the opponent's reaction time, which is especially critical for expert players [58]. These players often train their visual perception to better anticipate incoming strokes near the edges [59]. To practice placing the ball in hard-to-reach areas, such as the table's edge, players typically use a constraints-led approach [60].

In our study, we measured the ball's landing position as a key parameter. Specifically, the bounce distance $X$ refers to the distance from the side of the table to the bounce point, and the bounce distance $Y$ denotes the distance from the net, as illustrated in Fig. 4. By 'edge', we refer to the opposite side of the table from where the player is executing the stroke. For instance, if the player is hitting the forehand stroke from the right side, the ball should be directed towards the left side of the table. In our analysis, we assumed that both the $X$ and $Y$ values should be maximized for an optimal stroke. The method for obtaining these measurements involved using two cameras, with further details on the process discussed in subsequent sections.

#### 2) NET CLEARANCE

In a well-executed stroke, the net clearance is typically low but sufficient to avoid hitting the net. A good stroke should maintain a balance between minimizing the height of the ball over the net to make it harder for the opponent to attack, while still ensuring enough clearance to avoid the risk of an error [61]. Low net clearance keeps the ball's trajectory more direct, making it more difficult for the opponent to predict and return effectively, especially if combined with precise placement near the edges of the table.

#### 3) INSTANTANEOUS SPEED

The importance of speed in forehand strokes in table tennis lies in its ability to reduce the opponent's reaction time and increase the difficulty of returning the stroke. At high levels of play, players rely heavily upon quick reflexes, hence a fast stroke can prevent them from setting up for a strong counterattack [62].

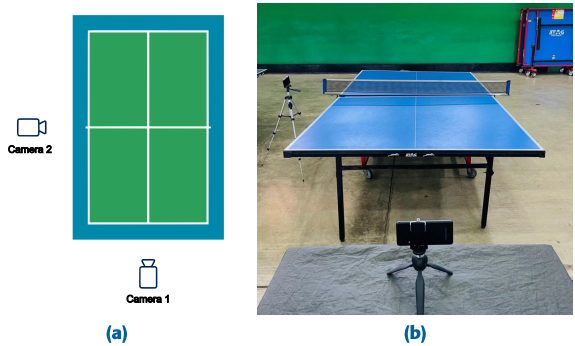

FIGURE 5. Camera setup for the data collection.

#### 4) HEIGHT RATIO
In the forehand technique of table tennis, it is widely recommended that the racket strike the ball at or near its maximum height during the ball's trajectory [63]. To quantify this, we introduced a height ratio parameter, which compares the actual contact point of the racket with the theoretical maximum height of the ball.

### C. DATA COLLECTION
#### 1) ENVIRONMENT FOR DATA COLLECTION
The data collection process was carried out at the Table Tennis courts at the gymnasium of the University of Peradeniya, Sri Lanka. To capture quality parameters pertaining to table tennis, two video cameras were placed during the data collection. The configuration of the test environment is given in Fig. 5.

Each individual was instructed to wear the wireless wearable device in the configuration described earlier and to perform $N_r = 50$ repetitions of a forehand stroke.

To analyze the quality of the forehand stroke in table tennis, we conducted an extensive data collection process involving $N_p = 18$ individuals. Each participant performed $N_r = 50$ forehand strokes. These strokes were recorded using two cameras at 60 fps, capturing the motion from different angles to ensure comprehensive data for quality parameter extraction for each stroke.

#### 2) DISTRIBUTION OF INDIVIDUALS
Raw accelerometric and gyroscopic data were collected through the wearable device from volunteers, with varying levels of expertise in table tennis. This group comprised beginners, players with considerable experience, and a few national players regarded as some of the best players in the country. This distribution was intentionally designed to approximate a fair distribution among participants, ensuring a balanced mix of skill levels. The details of the individuals are given in Table 3.

TABLE 3. Details of individuals.

| Symbol | Parameter | Value |
|---|---|---|
| $N_p$ | Number of individuals | 18 |
| $N_r$ | Number of realizations per person | 50 |
| $H$ | Height | 164 ± 4.23 cm |
| $M$ | Weight | 51 ± 5.22 kg |
| $a$ | Age | 25 ± 2.66 years |

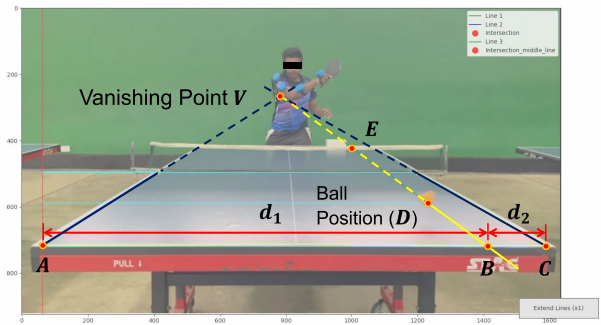

FIGURE 6. Perspective correction for the calculation of bounce distance X and Y.

### D. PARAMETER EXTRACTION
This section presents the methodology that was followed for the extraction of parameters using the collected video data.

#### 1) BOUNCE DISTANCE X AND Y
Images capturing each stroke's contact with the table were taken from camera 1, and the bounce distances for each stroke in the $X$ and $Y$ directions were measured for all $N_p = 18$ individuals. A Python script was developed to determine the ball's landing position ($X$ and $Y$) on the table from the images, accounting for perspective correction using the cross ratio theorem [64].

For this purpose, four points on the opposite side of the player on the table together with the position of the ball were marked after calculating the actual distance represented by a pixel. Then, the vanishing point ($V$) was obtained by extending the two lines created by the four points marked on the table edges as in Fig. 6.

Then, the bounce distance $X$ was calculated using:

$$p_1 \equiv X = W_{\text{table}} \times \frac{d_1}{d_1 + d_2}. \qquad (23)$$

Here, $d_1$ and $d_2$ are $AB$ and $BC$ pixel distances of the image, where point $B$ is the intersection of the line $AC$ and the extended line $DV$.

For the bounce distance $Y$, (24) was derived using cross ratio:

$$p_2 \equiv Y = \frac{(DE)(BV)}{(BE)(DV)} \times L_{\text{table}}. \qquad (24)$$

$DE$, $BE$, $DV$, and $BV$ are pixel distances where $D$ (the position of the ball), $E$ points are highlighted in Fig. 6.

**TABLE 4.** Standard dimensions for table tennis.

| Symbol | Parameter | Value |
|---|---|---|
| $L_{table}$ | Table length | 275.00 cm |
| $W_{table}$ | Table width | 152.50 cm |
| $h_{net}$ | Net height | 15.25 cm |
| $D_{ball}$ | Ball diameter | 4.00 cm |

#### 2) NET CLEARANCE

The videos from the two cameras were synchronized, and images were captured from the video recording of camera 1 when the ball was positioned directly above the net when observed from the camera 2 visual feed. If $h_{ball}$ is the elevation of the ball above the table, net clearance $n_c$ is calculated as:

$$p_3 \equiv n_c = h_{ball} - h_{net}. \quad (25)$$

#### 3) INSTANTANEOUS BALL SPEED IN THE Y DIRECTION

This parameter was defined to be the instantaneous speed of the ball above the net, approximated using a finite distance. The positions of two points of the ball's trajectory over the net within three frames of the video feed from camera 2 were noted. Given that the video was recorded at 60 frames per second, the speed of the ball was computed using the distance $d_{ball}$ between the aforementioned two points (assumed to be horizontal), and the time difference calculated via the frame rate is given by:

$$p_4 \equiv v_{ball,avg} = \frac{d_{ball}}{60 \text{ fps}} \times 3 \text{ frames}. \quad (26)$$

#### 4) HEIGHT RATIO

The height ratio $h_r$ of each stroke is dependent on two measurements of the elevation of the ball above the table during the time of the stroke. The elevations are at the following time instances:

- when the ball is released by the player ($h_0$), and
- when the ball makes contact with the racket ($h_1$).

These two measurements were obtained using the video feed from camera 1. It was assumed that the movement of the ball away from the vertical plane after the release was negligible so no perspective correction was necessary.

Then, the height ratio was calculated from:

$$p_5 \equiv h_r = 1 - \frac{h_1}{h_0}. \quad (27)$$

### IV. RESULTS AND DISCUSSION
#### A. OBTAINING THE PERFORMANCE SPACE
##### 1) DEFINITION OF COST FUNCTIONS

The definitions of the cost functions were based on domain expertise on table tennis performance supported by literature, and they are given in Fig. 7a. In addition, the standard dimensions for table tennis, which are given in Table 4, were also used in this task.

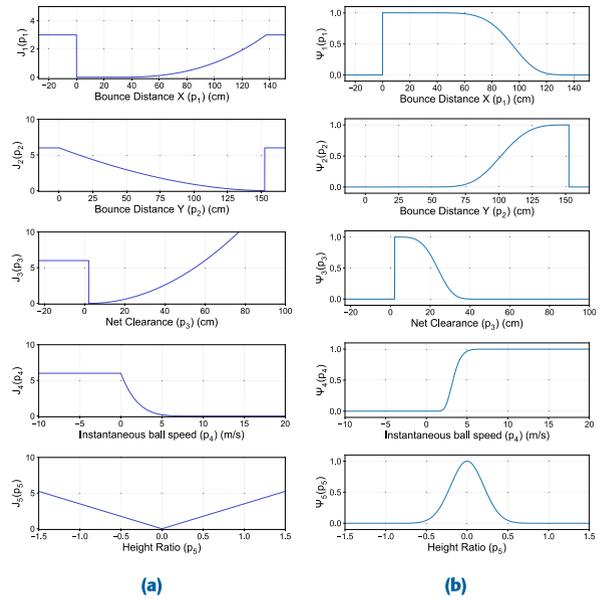

**FIGURE 7.** (a) Cost functions and (b) performance functions obtained through RBF mapping.

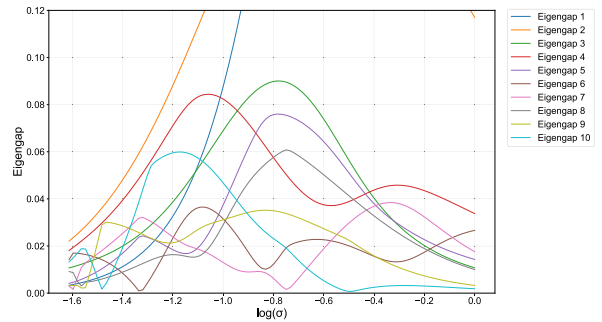

**FIGURE 8.** First 10 eigengaps for $N_p = 18$ individuals.

##### 2) NORMALIZATION THROUGH RBFS

The defined cost functions were then transformed into performance functions via RBFs as given in Fig. 7b.

#### B. CLUSTERING TO OBTAIN THE BENCHMARK CLUSTER

As the primary prerequisite to finding the benchmark cluster, clustering was utilized to identify any inherent partitions within the performance space $\mathbb{S}$ through spectral and $k$-means clustering.

##### 1) SPECTRAL CLUSTERING

The usage of spectral clustering on the performance space resulted in the eigengap vs. $\log(\sigma)$ curves presented in Fig. 8 for the first 10 eigengaps. This figure suggests that the ideal number of clusters for this scenario is $K_{sc} = 4$, as the curve for the fourth eigengap dominates and varies over a wider range of $\sigma$ values. It should be noted that this is concluded after excluding the first two eigengaps, which are trivial cases.

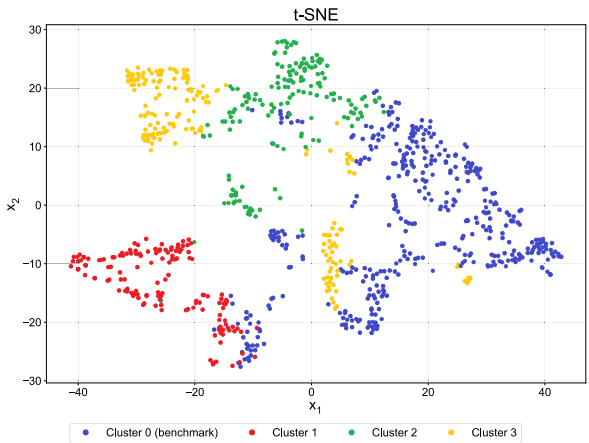

FIGURE 9. Using t-SNE to visualize clusters in 2D.

TABLE 5. Mean positions and euclidean distances for the obtained clusters.

| Cluster Index | Coordinates of mean position of the cluster | Euclidean distance |
|---|---|---|
| 0 | [0.9342, 0.8285, 0.8791, 0.7104, 0.4875] | 0.6284 |
| 1 | [0.2952, 0.4005, 0.0377, 0.2540, 0.5256] | 1.6012 |
| 2 | [0.2121, 0.8219, 0.9122, 0.6381, 0.6757] | 0.9468 |
| 3 | [0.3364, 0.1023, 0.9221, 0.4666, 0.5199] | 1.3294 |

### 2) K-MEANS CLUSTERING

The number of clusters $K_{sc} = 4$ established through spectral clustering was used to cluster the performance space $\mathbb{S}$ using $k$-means clustering. In order to obtain a visualization of these clusters, t-SNE [65] was used to dimensionally reduce the 5D space to a 2D space as shown in Fig. 9.

### C. CLUSTER MEAN DISTANCES FROM THE ORIGIN OF IDEAL PERFORMANCE O′

After selecting the 4 primary clusters, the centroid of each cluster was calculated using (19). Subsequently, the distance from the centroid of each cluster to the origin of ideal performance $O' \equiv (1, 1, 1, 1, 1)$ in the performance space was calculated using (20). The results are summarized in Table 5.

With the results obtained in Table 5 for the mean Euclidean distances, cluster $q = 0$ was identified as the benchmark cluster with the lowest distance of $d_0 = 0.6284$ from $O'$ considering (21).

### D. DISTRIBUTION OF INDIVIDUALS IN THE BENCHMARK CLUSTER

It is essential to identify how the data points from each individual are categorized into each of the main clusters. Fig. 10 exhibits the cluster composition for each individual. It can be identified that the benchmark cluster includes data points from all individuals, emphasizing that it captures the well-performed forehand strokes of the players irrespective of their level of expertise.

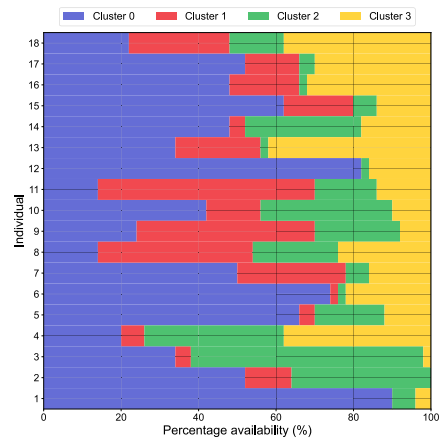

FIGURE 10. Distribution of each individual in each cluster.

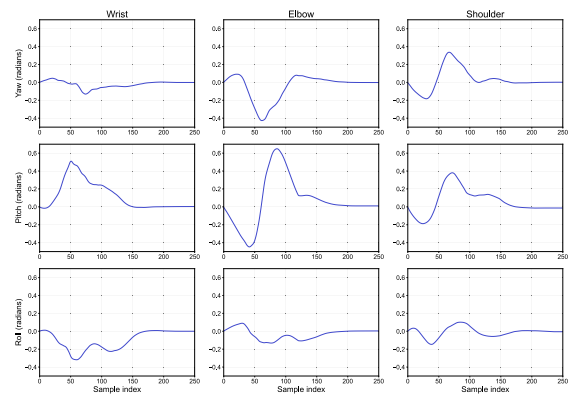

FIGURE 11. Mean Euler angle variation of benchmark cluster (cluster 0).

### E. MEAN EULER SIGNAL AND REALIZATIONS IN THE BENCHMARK

The mean Euler signal of each cluster was obtained using (22). Fig. 11 shows the variation in the Euler angles of wrist, elbow, and shoulder for the strokes grouped within the benchmark cluster. At the wrist, minimal variation is observed in the yaw and roll angles compared to the pitch angle. A positive pitch angle indicates a subtle push-like motion from the wrist, contributing to generating pace on the ball via the wrist. In the elbow joint, there is a slight negative rotation around the $y$-axis, followed by a smooth transition into positive rotation, indicating an effective transmission of power and a natural flow of the upper body during the stroke.

### F. COMPARISON OF THE MEAN EULER SIGNAL OF EACH CLUSTER

When analyzing the mean Euler signals for each cluster in Fig. 12, starting with yaw and roll angles of the wrist, the benchmark cluster shows steady yaw with minimal deviation, as previously noted. In contrast, the other clusters exhibit higher fluctuations in yaw, possibly reflecting variation in wrist rotations around the $z$-axis within those grouped strokes. Regarding the roll angle, cluster 2 stands out with significantly higher rotations, suggesting extensive palm rotations around the $x$-axis, while the other two clusters closely follow the same pattern in reduced magnitude. The deviations in roll angle from the benchmark cluster indicate that, while the strokes aimed to minimize the net clearance, the other cluster strokes appeared to focus on clearing the net by adjusting the racket angle to facilitate an upward trajectory of the projectile.

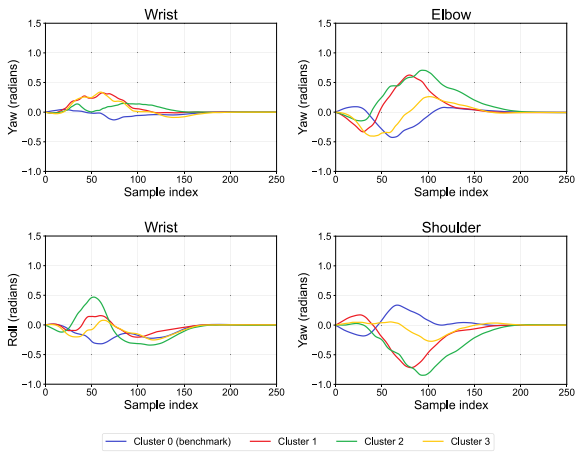

FIGURE 12. Mean Euler angle variations of the 4 clusters.

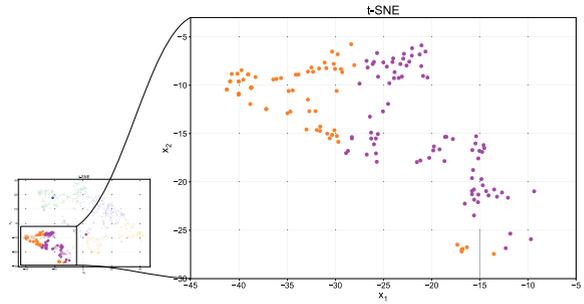

FIGURE 13. Visualization of sub-clusters of cluster 1 using t-SNE.

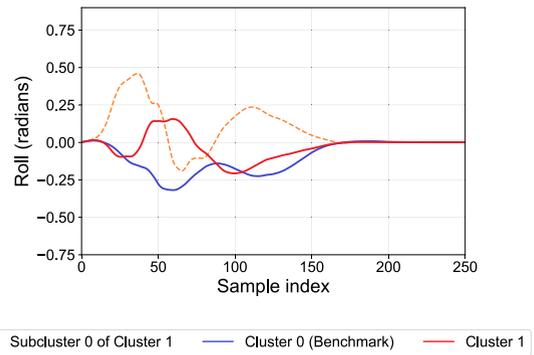

FIGURE 14. Comparison between the benchmark mean signal, mean signal of a cluster, and the mean signal of a sub-cluster of cluster 1 considering wrist roll angle.

Examining the elbow angle during a forehand stroke, the benchmark cluster displays a slight positive peak followed by a negative peak, indicating a small rotation around the $z$-axis toward the ball, with the rotation reversing after contact with the racket. However, in the other clusters, the pattern is reversed, suggesting that corrections may be needed in the technique. When the shoulder yaw angle is considered also, a similar pattern emerges, with the other clusters exhibiting the exact opposite movement compared to the benchmark cluster. In summary, the benchmark cluster strokes involved the elbow and shoulder pushing forward, followed by a retraction after ball contact. In contrast, the other clusters suggest a technique where the racket starts positioned further behind the body and moves forward to meet the ball's trajectory, potentially requiring a technique adjustment.

### G. ANALYSIS OF SUB-CLUSTERS

By examining a divergent cluster more closely, we can further divide it into sub-clusters to gain deeper insights into the strokes within that cluster. If we consider the cluster that was the most distant in Euclidean space from the origin of the parameter space (cluster 1) and apply the previously discussed clustering methodology, cluster 1 was divided into further two clusters. Fig. 13 depicts the t-SNE plot of the obtained clusters. As shown in Fig. 14, the mean roll angle of the wrist joint in one of those sub-clusters (from cluster 2) deviates further from the benchmark mean, highlighting the lack of control in the wrist movement within this sub-cluster. This indicates that by utilizing the proposed framework, it is also possible to zoom in on any cluster and even analyze individual strokes to identify specific movement patterns or issues related to those executions.

### V. CONCLUSION

This study presents an approach to benchmarking psychomotor skills, with a specific application to table tennis forehand strokes. Motion data captured through IMU sensors revealed four distinct clusters within the mapped performance space, each characterized by unique biomechanical patterns. A benchmark cluster was established based on its proximity to the origin in the performance space, signifying optimal performance. Deviations from this benchmark were analyzed using Euler angles to quantify differences in wrist, elbow, and shoulder movements. Specifically, the benchmark cluster exhibited minimal wrist variation, a subtle push-like motion, and smooth elbow transitions. In contrast, divergent clusters demonstrated larger perturbations of the wrist and elbow, indicative of suboptimal movement patterns.

This research introduced a data-driven framework that combines wearable sensor technology with unsupervised clustering techniques. By mapping biomechanical data to a performance space, we can objectively assess and compare performance levels. This methodology offers an effective technique for identifying strengths and weaknesses in individual performance, enabling the development of targeted training interventions. Furthermore, the analysis of motion space deviations provides valuable insights into the biomechanical factors underlying performance differences.

## VI. FUTURE WORK AND RECOMMENDATIONS

The framework outlined in this study can be applied to evaluate tasks and design coaching or learning experiments aimed at developing specific skill sets to enhance performance in selected psychomotor tasks. In order to adapt this framework for assessing a different psychomotor task, relevant performance parameters should be defined to align with the optimal execution of that task. By constructing a corresponding performance space and employing the same methodology, valuable insights into motion dynamics can be obtained. A larger sample population would yield more precise and informative clusters in the performance space which would aid in the analysis of corresponding movement kinematics in the motion space. Utilizing this framework presented in the paper, more refined models can be developed to analyze performance outcomes, ultimately leading to more personalized training programs in a wide variety of applications including medicine, rehabilitation, and industry. This adaptability extends to activities requiring fine motor skills or more complex full-body movements, as the wearable device allows for various subsets of sensors to be selected depending on the task. Furthermore, integrating additional sensors such as EMG with IMU data will facilitate a multi-modal approach, offering deeper insights into muscle activation patterns and further enhancing the scope of psychomotor skill assessments and improvement.

## ACKNOWLEDGMENT
The authors would like to thank the Faculty of Engineering, University of Peradeniya, Sri Lanka, especially the Department of Electrical and Electronic Engineering (DEEE), for their unwavering support during the course of their research project. They would also like to thank Dr. Pramila Gamage, the Head of the Manufacturing and Industrial Engineering Department (DMIE) for her invaluable guidance toward the success of this project.

Next, they acknowledge the contributions of the volunteers who participated in the data collection process, especially Shiham Firdous, Tharindi Rathnayake, and Bhashini Jayathunge, undergraduates from the DEEE for their assistance in the quality parameter extraction process using visual cameras.

Moreover, they would like to express their gratitude to the University Research Council (URC), and the University of Peradeniya for the encouragement and support for multidisciplinary research.

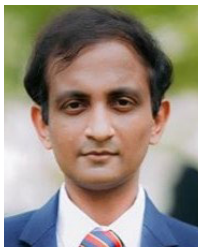

**MAHELA PANDUKABHAYA** (Member, IEEE) received the B.Sc. degree in electrical and electronic engineering from the University of Peradeniya, Peradeniya, Sri Lanka, in 2024.

He is currently a Teaching Assistant with the Department of Electrical and Electronic Engineering, Faculty of Engineering, University of Peradeniya, while actively involving in research related to signal processing, artificial intelligence, and biomedical engineering, focusing on biomedical wearables in particular.

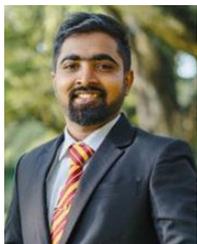

**THARAKA FONSEKA** received the B.Sc. (Eng.) degree in electrical and electronic engineering from the University of Peradeniya, Peradeniya, Sri Lanka, in 2024.

He is currently a Research Assistant with the Multidisciplinary AI Research Centre, University of Peradeniya. He has published papers in IEEE conferences and journals. His research interests include applications of artificial intelligence in human behavior, pandemic analysis, biomedical wearables, pandemic analysis, signal processing, machine learning, statistics, AI, and telecommunications.

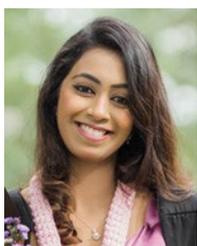

**MADHUMINI KULATHUNGE** (Member, IEEE) received the B.Sc. (Eng.) degree in electrical and electronic engineering from the University of Peradeniya, Peradeniya, Sri Lanka, in 2024.

She is currently a Teaching Assistant with the Department of Engineering Mathematics, University of Peradeniya. In addition to her teaching duties, she actively participates in research on biomedical wearables utilizing inertial measurement unit (IMU) sensor systems. Her research has been published at IEEE conferences. Her research interests include machine learning, artificial intelligence, and signal processing.

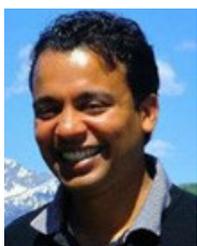

**ROSHAN GODALIYADDA** (Senior Member, IEEE) received the B.Sc. (Eng.) degree in electrical and electronic engineering from the University of Peradeniya, Peradeniya, Sri Lanka, in 2005, and the Ph.D. degree from the National University of Singapore, Singapore, in 2011.

He is currently a Professor with the Department of Electrical and Electronic Engineering, Faculty of Engineering, University of Peradeniya. He has numerous publications in leading international journals and conferences, such as IEEE TRANSACTIONS ON REMOTE SENSING AND GEOSCIENCE, *Applied Energy*, IEEE TRANSACTIONS ON SMART GRID, *Journal of Food Engineering*, *International Journal of Electrical Power and Energy Systems*, IEEE TRANSACTIONS ON MEASUREMENT AND INSTRUMENTATION, IEEE JOURNAL OF SELECTED TOPICS IN APPLIED EARTH OBSERVATIONS AND REMOTE SENSING, *PLoS ONE*, *Sensors*, IEEE ACCESS, ICIP, WCNC, and IGARSS to name a few. His current research interests include signal and image processing, computer vision, machine learning, smart grids, renewable energy integration, biomedical engineering, computational epidemiology, remote sensing, and spectral imaging.

Dr. Godaliyadda was a recipient of multiple grants through the National Science Foundation (NSF), Sri Lanka, and International Development Research Center (IDRC), Canada. He was also a recipient of multiple best paper awards from international conferences for his work. He received Sri Lanka President's Award for Scientific Research in 2016, 2017, 2019, and 2020.

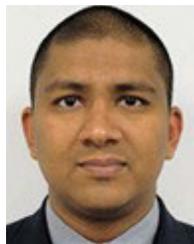

**PARAKRAMA EKANAYAKE** (Senior Member, IEEE) received the B.Sc. (Eng.) degree in electrical and electronic engineering from the University of Peradeniya, Peradeniya, Sri Lanka, in 2006, and the Ph.D. degree from Texas Tech University, Lubbock, TX, USA, in 2011.

He is currently a Professor with the University of Peradeniya. His previous works have been published in IEEE TRANSACTIONS ON GEOSCIENCE AND REMOTE SENSING and several other IEEE-GRSS conferences, including WHISPERS and IGARSS. He also has multiple publications in many IEEE TRANSACTIONS, Elsevier, and IET journals. His current research interests include applications of signal processing and system modeling in remote sensing, hyperspectral imaging, and smart grids.

Dr. Ekanayake was a recipient of Sri Lanka President's Award for Scientific Publications, in 2018 and 2019. He has obtained several grants through the National Science Foundation (NSF) for research projects. He has been awarded several best paper awards at international conferences.

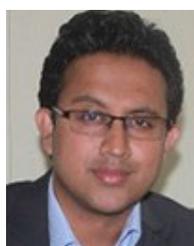

**CHANAKA SENANAYAKE** received the B.Sc. (Eng.) degree in production engineering from the University of Peradeniya, Peradeniya, Sri Lanka, in 2005, and the Ph.D. degree from the National University of Singapore, Singapore, in 2012.

He was a Visiting Research Scholar with the Division of Production Systems, Politecnico di Milano, Italy, from October 2013 to December 2013, and a Postdoctoral Research Associate with the Department of Industrial, Manufacturing and Systems Engineering, Texas Tech University, Lubbock, TX, USA, from September 2019 to January 2021. He is currently a Senior Lecturer with the Department of Manufacturing and Industrial Engineering, Faculty of Engineering, University of Peradeniya. He has published in leading international journals and conferences, including *International Journal of Production Economics*, *European Journal of Mechanics, A/Solids*, *Engineering Failure Analysis*, *Flexible Services and Manufacturing Journal*, and *Computers and Industrial Engineering*. His current research interests include performance evaluation, stochastic modeling of manufacturing and service systems, operations management, systems simulation, and healthcare delivery.

Dr. Senanayake was a recipient of the 2021 President's Excellence in Engaged Scholarship awarded by Texas Tech University for his community engagement efforts during the global pandemic. He was the General Chair of the First National Symposium on Manufacturing and Industrial Engineering, Sri Lanka, in 2016. He was also a Programme and the Editorial Chair of the Production and Operations Management Society International Conference (POMS), Sri Lanka, in 2018.

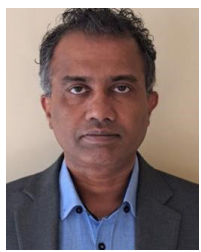

**VIJITHA HERATH** (Senior Member, IEEE) received the B.Sc. Eng. degree (Hons.) in electrical and electronic engineering from the University of Peradeniya, Peradeniya, Sri Lanka, in 1998, the M.Sc. degree in electrical and computer engineering from the University of Miami, Coral Gables, FL, USA, in 2002, with the award of academic merit, and the Ph.D. degree in electrical engineering from the University of Paderborn, Paderborn, Germany, in 2009.

He is currently a Professor with the Department of Electrical and Electronic Engineering, University of Peradeniya. His current research interests include remote sensing, multispectral imaging, AI and machine learning, and light communication.

Dr. Herath is a Senior Member of Optica and a member of the Institution of Engineers, Sri Lanka. He was a recipient of Sri Lanka President's Award for Scientific Research, in 2013 and 2020. He received the paper awards at the ICTer 2017 Conference and MERCon 2023 Conference. He was the General Chair of the IEEE International Conference on Industrial and Information Systems (ICIIS), Kandy, Sri Lanka, in 2013.

○ ○ ○